\documentclass[
  reprint,
  superscriptaddress, 
  amsmath,amssymb,
  aps,
  prl
]{revtex4-2}

\usepackage{graphicx}
\usepackage{dcolumn}
\usepackage{bm}
\usepackage{dsfont}
\usepackage{tikz}
\usepackage{indentfirst}
\usepackage{multirow}

\usepackage[colorlinks,citecolor=blue,linkcolor=red,urlcolor=magenta]{hyperref}
\newcommand{\remove}[1]{}

\newcommand{\supplementarysection}{%
  \setcounter{figure}{0}
  \let\oldthefigure\thefigure
  \renewcommand{\thefigure}{S\oldthefigure}
  \setcounter{section}{0}
  \let\oldthesection\thesection
  \renewcommand{\thesection}{S\oldthesection}
  \setcounter{equation}{0}
  \let\oldtheequation\theequation
  \renewcommand{\theequation}{S\oldtheequation}
  \setcounter{table}{0}
  \let\oldthetable\thetable
  \renewcommand{\thetable}{S\oldthetable}
}

\newcommand{\bi}{\begin{itemize}}
\newcommand{\ei}{\end{itemize}}
\newcommand{\be}{\begin{enumerate}}
\newcommand{\ee}{\end{enumerate}}

\newenvironment{dfn}{{\vspace*{1ex} \noindent \bf Definition }}{\vspace*{1ex}}

\newcommand{\nn}{\nonumber}  %

\newcommand{\Eq}[1]{Eq.~(\ref{#1})}

\newcommand{\ket}[1]{\left| #1 \right>} 
\newcommand{\trace}[1]{\mathrm{tr}{\left(#1\right)}}
	\newcommand{\bra}[1]{\left< #1 \right|} 
	
	\newcommand{\beq}{\begin{eqnarray}}
	\newcommand{\eeq}{\end{eqnarray}}
	\newcommand{\bea}{\begin{eqnarray}\begin{aligned}}
	\newcommand{\eea}{\end{aligned}\end{eqnarray}}

\definecolor{forestgreen}{RGB}{34,139,34}

\begin{document}

\title{Primary charge-$4e$ superconductivity from doping a featureless Mott insulator}

\author{Zhi-Qiang Gao}
\thanks{These authors contributed equally.}
\affiliation{Department of Physics, University of California, Berkeley, California 94720, USA}
\author{Yan-Qi Wang}
\thanks{These authors contributed equally.}
\affiliation{Department of Physics and Joint Quantum Institute, University of Maryland,
College Park, Maryland 20742, USA}
\author{Ya-Hui Zhang}
\email{yzhan566@jhu.edu}
\affiliation{Department of Physics and Astronomy, Johns Hopkins University, Baltimore, Maryland 21218, USA}
\author{Hui Yang}
\email{huiyang.physics@gmail.com}
\affiliation{Department of Physics and Astronomy, University of Pittsburgh, PA 15213, USA}
\affiliation{Department of Physics and Astronomy, Johns Hopkins University, Baltimore, Maryland 21218, USA}

\begin{abstract}
Superconductivity is usually understood as a phase in which charge-$2e$ Cooper pairs are condensed.  Charge-$4e$ superconductivity has largely been discussed as a vestigial order at finite temperature emerging from charge-$2e$ states. Primary charge-$4e$ superconducting phases at zero temperature remain scarce in both experiments and microscopic models. Here we argue that a doped featureless Mott insulator with $SU(4)$ symmetry provides a natural platform for primary charge-$4e$ superconductivity, based on perturbative renormalization group arguments and group theoretic considerations. As a concrete realization, we construct a bilayer Hubbard model with tunable onsite $SU(4)$ and $Sp(4)$ symmetries that exhibits a featureless Mott insulating phase at half filling. Its low energy physics is captured by a generalized ESD model, featuring an effective Hamiltonian that is purely kinetic within the constrained Hilbert space. Using density matrix renormalization group (DMRG) simulations, we find a primary charge-$4e$ superconducting phase in the $SU(4)$ ESD model and a conventional primary charge-$2e$ phase in the $Sp(4)$ case. We further characterize the corresponding normal states and discuss the resulting finite temperature phase diagram.
\end{abstract}

\maketitle

\textit{Introduction.---} Superconductivity (SC) is, by default, understood as a phase in which charge-$2e$ Cooper pairs are condensed. In principle, however, the elementary condensed object can be a bound state of more than two electrons. The simplest example is a charge-$4e$ SC, in which quartets of electrons are condensed while charge-$2e$ Cooper pairs are gapped. Here, the quartet denotes a bound state of four electrons and serves as the system's order parameter. Charge-$4e$ SC has been studied primarily in the context of pair density wave states~\cite{Berg2009,Radzihovsky2009,Radzihovsky2011,Wu20242} and multicomponent superconductors~\cite{Vidal2002,Babaev2004,Radzihovsky2004,Radzihovsky2008,Herland2010,Jian2021,Fernandes2021,Grinenko2021,Liu2023,Babaev2024,Zeng2024,Volovik2024,Soldini2024,Dai2024,Chirolli2024,Hecker2024,samoilenka2025,zou2025,Song2025}, as a vestigial order at finite temperature. In contrast, platforms for \emph{primary} charge-$4e$ SC at {\it zero temperature} are less discussed, with theoretical proposals on low-dimensional fermionic systems~\cite{Wu2005,Lecheminant2005,Capponi2007,Capponi2008,Roux2008,Gnezdilov2022} and their recent generalizations to moir\'e systems~\cite{Zhang2020,Khalaf2022,Wu2024}, doped $SU(4)$ chiral spin liquids~\cite{Zhang20254e,Pichler2025}, and its wavefunctions~\cite{Hu4e,Gao2025} and possible topological properties~\cite{Wang2016me,Gao2025,shi2025nonabelian,shi2026}. To our knowledge, a concrete lattice model that robustly hosts a primary charge-$4e$ SC ground state in two dimensions is still lacking. On the experimental side, quartet formation~\cite{Gladchenko2008,Cohen2018,Huang2022,Ciaccia2024} and vestigial charge-$4e$ SC~\cite{Ge2024} have been reported, but an unambiguous realization of a primary charge-$4e$ SC phase has yet to be identified.

Conceptually, from the perspective of perturbative renormalization group analysis around a Fermi liquid fixed point, a charge-$4e$ SC instability is always less relevant than that of conventional charge-$2e$, simply by power counting in the number of electron operators entering the order parameter. Within a weak coupling framework with a sizable Fermi surface, it is therefore extremely difficult to obtain a primary charge-$4e$ SC phase, and dynamical mechanisms that suppress charge-$2e$ instabilities become essential. The most natural consideration is to drive the system far away from the Fermi liquid fixed point into a strongly correlated regime, where naive power counting arguments cease to apply. Another is to realize a small or strongly reconstructed Fermi surface, such that conventional charge-$2e$ Cooper pairing is severely phase-space constrained while charge-$4e$ quartets can still delocalize efficiently. A complementary possibility is to impose strong frustration on single electron or Cooper pair motion while allowing appropriate multi-electron bound states to propagate coherently, so that the kinetic energy gain of quartets dominates. All of these types of dynamical constraints are saturated in the recently proposed effective model composed of empty, singlon, and doublon states, dubbed the \emph{ESD model}, which is a low energy effective theory of bilayer nickelates La$_3$Ni$_2$O$_7$ and an example of kinetic energy driven SC~\cite{Oh2024,Yang2025,Oh2025,ESDreview}. Although the primary SC phase in the original ESD model is still charge-$2e$, we expect suitable generalizations of the ESD construction to give rise to primary charge-$4e$ (or even charge-$2ne$) SC phases.

On the other hand, large internal symmetry group imposes kinematic constraints on the charge of singlet SC instabilities. In classifying and constraining the charged low energy excitations, the concept of group center has been widely used across modern quantum many-body and field theory. Here, the center $Z(G)$ is the maximal subgroup of $G$ that commutes with all elements of $G$. Inspired by the study of quark confinement and hadron formation~\cite{Zee,Greensite2003}, we show, based on group theory, that the lowest charge of the primary singlet SC is rigorously constrained by the nontrivial center of the internal continuous symmetries of the system~\cite{supp}. In particular, $SU(4)$ symmetry enforces the lowest charged singlet SC instability to have charge-$4e$. As a comparison, $Sp(4)$ admits singlet SC instability for both charge-$2e$ and charge-$4e$. This suggests that, with $Sp(4)$ symmetry, a charge-$4e$ SC phase is typically vestigial, as its order parameter always contains a composite of two charge-$2e$ ones, $\Delta_{4e}\sim\Delta_{2e}\Delta_{2e}$. With $SU(4)$ symmetry, however, the absence of a singlet charge-$2e$ order parameter ensures a singlet charge-$4e$ SC phase to be primary and persist to zero temperature if the $SU(4)$ symmetry is not broken. We term this kinematic constraint from symmetry group center as {\it center enforcement mechanism}. Simply connected Lie groups enforcing primary SC with charge $Q>2e$ are summarized in Table~\ref{tab:1}. 

\begin{table}[!htbp]
\begin{tabular}{ccccc}
\hline\hline
Group $G$~ & ~$\mathrm{Spin}(4n\!+\!2)$~ & ~$E_6$~ & ~$SU(2n)$~ & ~$SU(n)$, $n$ odd \\
Charge $Q$~ & $4e$ & $6e$ & $2ne$ & $2ne$\\
\hline\hline
\end{tabular}
\caption{Primary singlet charge-$Q$ ($Q>2e$) SC enforced by the nontrivial center of Lie group $G$. In particular, primary charge-$4e$ SC is favored by $SU(4)=\mathrm{Spin}(6)$ symmetry.}\label{tab:1}
\end{table}

Taken together, the dynamical constraints pointing to a strongly correlated regime and the center enforcement mechanism favoring enlarged internal symmetry suggest a doped featureless Mott insulator~\cite{Kimchi2013} with $SU(4)$ symmetry. These provide a natural platform for a primary charge-$4e$ SC phase. In this work, we propose a concrete lattice construction based on an $SU(4)$ symmetric bilayer Hubbard model whose low energy physics is captured by a generalized ESD model. For comparison, we also explicitly break the $SU(4)$ symmetry down to $Sp(4)$, and show that the resulting theory supports a conventional primary charge-$2e$ SC instead. To the best of our knowledge, this provides the first concrete two dimensional lattice construction of a primary charge-$4e$ SC phase at zero temperature.

\textit{The model.---} We consider the following bilayer Hubbard model on the square lattice as the parent Hamiltonian for the generalized ESD model:
\beq
H&=&-t\sum_{\left<\mathbf{i}\mathbf{j}\right>}c_{l,\sigma}^\dagger(\mathbf{i})c_{l,\sigma}(\mathbf{j})+\frac{u}{2}\sum_{\mathbf{i},l}n^2_{l}(\mathbf{i})+v\sum_{i}n_+({\bf i})n_-({\bf i})\nn\\
&\quad & +J_L\sum_{\mathbf{i}}L^{ab}_+(\mathbf{i})L^{ab}_-(\mathbf{i})+J_V\sum_{\mathbf{i}}V^a_+(\mathbf{i})V^a_-(\mathbf{i}).\label{eq:bH}
\eeq
Here, $c_{l,\sigma}^\dagger(\mathbf{i})$ creates an electron in layer $l = \pm$ along with flavor index $\sigma = 1,2,3,4$. The positive $t$ and $u$ describes intralayer hopping and Hubbard interaction, respectively.  The interlayer density-density interaction term associated to $v$ is symmetric under $SU(4)$ and $Sp(4)$. In the second line, $J_L\ge J_V\ge 0$ describes the interlayer $Sp(4)$ Heisenberg interaction which at proper fillings favors singlet on each rung connecting sites $\mathbf{i}$ in two layers. The electrons in each layer transform in the fundamental irreducible representation (irrep) $\mathbf{4}$ of $Sp(4)$. $L_\pm^{ab}(\mathbf{i})=(1/2)c^\dagger_{\pm,\sigma}(\mathbf{i})\Gamma^{ab}_{\sigma\sigma^\prime}c_{\pm,\sigma^\prime}(\mathbf{i})$ are $Sp(4)$ generators, while $V_\pm^a(\mathbf{i})=(1/2)c^\dagger_{\pm,\sigma}(\mathbf{i})\Gamma^{a}_{\sigma\sigma^\prime}c_{\pm,\sigma^\prime}(\mathbf{i})$ transform under the $\mathbf{5}$ vector irrep. They together form the generator set of $SU(4)$~\cite{Wu2003,Wu2005,Gao2020,supp}. Here $\Gamma^{a=1,2,3,4,5}$ are 4-by-4 gamma matrices with $\Gamma^{ab}=-(i/2)[\Gamma^a,\Gamma^b]$~\cite{supp}. When $J_V=J_L$, model \Eq{eq:bH} restores $SU(4)\times (U(1)_+\times U(1)_-/\mathbb{Z}_2)$ symmetry;  for general $J_V \neq J_L$, it has a symmetry of $ Sp(4)\times (U(1)_+\times U(1)_-/\mathbb{Z}_2)$, where the two $U(1)$ factors are corresponding to the charge conservations in each separate layer.

\begin{figure}[h]
    \centering
\includegraphics[width=1.\linewidth]{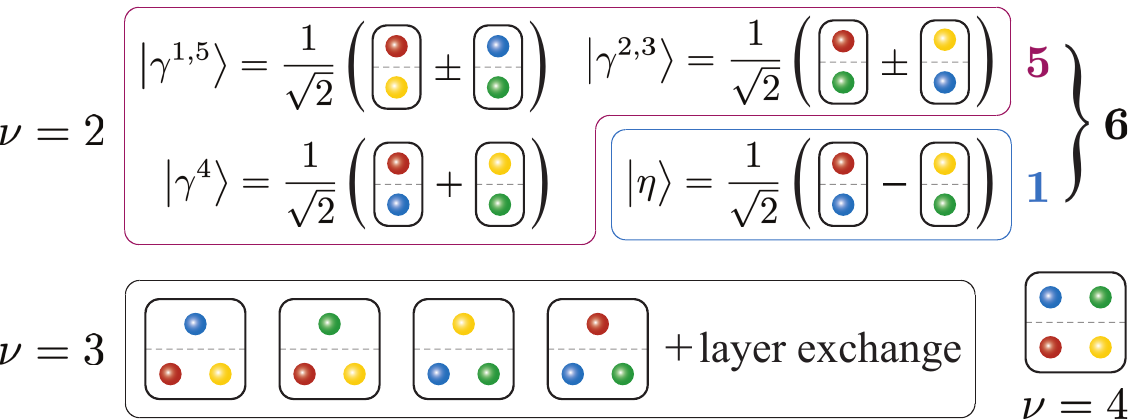}
\caption{Illustration of the low energy Hilbert space. The four colors denote the four flavors of electrons, and two layers are divided by the grey dashed line. In the $\nu=2$ sector, the $\mathbf{6}$ vector irrep of $SU(4)$ is decomposed into the $\mathbf{5}$ vector $\ket{\gamma^a(\mathbf{i})}$ and the $\mathbf{1}$ singlet $\ket{\eta(\mathbf{i})}$ under $Sp(4)$. For both $SU(4)$ and $Sp(4)$, in $\nu=3$ sector there are $4+4=8$ states related via layer exchange, and in $\nu=4$ sector there is a unique singlet state. }
\label{fig:state}
\end{figure}

The local Hilbert space of model \Eq{eq:bH} is defined on each rung connecting the two layers. In the Mott limit $u\gg J_{L,V}\gg t$, the low energy Hilbert space contains the ground states on rungs at integer fillings $\nu$. At half-filling $\nu=4$ with four electrons per rung, or two electrons per site per layer, the ground state on each rung is a singlet $\ket{q(\mathbf{i})}$ for both $SU(4)$ and $Sp(4)$ cases,  suggesting a featureless Mott insulator state. At $\nu=3$, there are $4+4=8$ states related by the layer exchange symmetry. At quarter filling $\nu=2$ with one electron per site per layer, the rung ground states for $SU(4)$ and $Sp(4)$ are different. It is crucial to note that, the $\mathbb{Z}_4$ center of the $SU(4)$ symmetry does not admit a singlet state at this filling, and the rung ground states are six-fold degenerate, transformed under the $\mathbf{6}$ vector irrep. When $SU(4)$ is explicitly broken to $Sp(4)$, the $\mathbf{6}$ vector is decomposed into a $\mathbf{5}$ vector $\ket{\gamma^a(\mathbf{i})}=-(i/2)c_{+,\sigma}^\dagger(\mathbf{i})(R\Gamma^a)_{\sigma\sigma^\prime}c^\dagger_{-,\sigma^\prime}(\mathbf{i})\ket{0}$ plus a singlet $\ket{\eta(\mathbf{i})}=(1/2)c_{+,\sigma}^\dagger(\mathbf{i})R_{\sigma\sigma^\prime}c^\dagger_{-,\sigma^\prime}(\mathbf{i})\ket{0}$~\cite{Yang1989}, as illustrated in Fig.~[\ref{fig:state}]. Here $R=\Gamma^1\Gamma^3$ is the charge conjugation matrix~\cite{supp}. The energy gap between $\ket{\gamma^a(\mathbf{i})}$ and $\ket{\eta(\mathbf{i})}$ is proportional to $J_L-J_V$, where $\ket{\eta(\mathbf{i})}$ becomes the unique ground state.

We derive the effective model of \Eq{eq:bH} by projecting it to the low energy subspace. The projection operator ${\mathcal P}$ associated with that low energy sector is defined as,
\beq
\mathcal{P}c_{l,\sigma}(\mathbf{i})\mathcal{P}&=&r_1\ket{\eta(\mathbf{i})}\bra{f_{l,\sigma}(\mathbf{i})}+r_5 \Gamma^a_{\sigma\sigma^\prime}\ket{\gamma^a(\mathbf{i})}\bra{f_{l,\sigma^\prime}(\mathbf{i})}\nn\\
&&+\ket{f_{-l,\sigma}(\mathbf{i})}\bra{q(\mathbf{i})},\label{eq:ele}
\eeq
where $r_1$ and $r_5$ are constants depending on microscopic parameter $J_L/J_V$, satisfying $r_1\!=\!r_5\!=\!2/\sqrt{3}$ when the $Sp(4)$ symmetry is restored to $SU(4)$. Here $\Gamma^a$ are gamma matrices defined before~\cite{supp}. Once projected to the low energy Hilbert space, the model \Eq{eq:bH} becomes the generalized ESD model,
\beq
H_\mathrm{ESD}&=&-t\sum_{\left<\mathbf{i}\mathbf{j}\right>}\mathcal{P}c_{l,\sigma}^\dagger(\mathbf{i})c_{l,\sigma}(\mathbf{j})\mathcal{P}+\sum_\mathbf{i}\epsilon \big(n_q(\mathbf{i})+n_\eta(\mathbf{i})\big)\nn\\
&&+\sum_\mathbf{i}(\epsilon+\delta)n_\gamma(\mathbf{i}),\label{eq:esd}
\eeq
where $n_q(\mathbf{i})=\ket{q(\mathbf{i})}\bra{q(\mathbf{i})}$, $n_\eta(\mathbf{i})=\ket{\eta(\mathbf{i})}\bra{\eta(\mathbf{i})}$, and $n_\gamma(\mathbf{i})=\sum_a\ket{\gamma^a(\mathbf{i})}\bra{\gamma^a(\mathbf{i})}$ are density operators. Coefficients $\epsilon$ and $\delta$ depend on the microscopic details, where $\epsilon$ characterizes the net interaction from the $v$ term and $J_{L,V}$ terms, while $\delta=0$ for $SU(4)$ and $\delta>0$ for $Sp(4)$ such that $\ket{\eta(\mathbf{i})}$ and $\ket{\gamma^a(\mathbf{i})}$ become degenerate when $Sp(4)$ is restored to $SU(4)$.

From doping holes to the featureless Mott insulator at $\nu=4$, the onset of primary charge-$2e$ SC in the $Sp(4)$ ESD model is similar to that in $SU(2)$ ESD model~\cite{ESDreview}, where the condensed singlet Cooper pair is $h^6(\mathbf{i})=\ket{q(\mathbf{i})}\bra{\eta(\mathbf{i})}$. However, for $SU(4)$, it is not obvious that the model \Eq{eq:esd} exhibits primary charge-$4e$ SC. In particular, when $\epsilon\ll 0$, {\it i.e.}, in the presence of a strong interlayer attraction, the Cooper pair $h^6(\mathbf{i})$ together with $h^a(\mathbf{i})=\ket{q(\mathbf{i})}\bra{\gamma^a(\mathbf{i})}$ should condense and spontaneously break $SU(4)$ to $Sp(4)$, leading to a flavor-polarized charge-$2e$ SC phase instead of the desired singlet charge-$4e$ SC phase. To understand the emergence of primary charge-$4e$ SC, we derive the effective theory of $SU(4)$ Cooper pairs $h^a(\mathbf{i})$ via a standard $t/|\epsilon|$-expansion starting from the $\epsilon\ll 0$ limit~\cite{supp}, where we view $\bigotimes_\mathbf{i}\ket{q(\mathbf{i})}$ as the vacuum and dope dilute density of Cooper pairs. The effective Hamiltonian reads
\beq
H_h&=&\frac{t^4}{64|\epsilon |^3}\sum_{\left<\mathbf{i}\mathbf{j}\right>} n_q(\mathbf{r}_{\left<\mathbf{i}\mathbf{j}\right>})\big(S^{ab}(\mathbf{i})S^{ab}(\mathbf{j})-3n_h(\mathbf{i})n_h(\mathbf{j})\big)\nn\\
&&-\frac{t^2}{2|\epsilon |}\sum_{\left<\mathbf{i}\mathbf{j}\right>}h^{a\dagger}(\mathbf{i})h^a(\mathbf{j}),\label{eq:h}
\eeq
where $S^{ab}(\mathbf{i})=i\big(h^{a\dagger}(\mathbf{i})h^b(\mathbf{i})-h^{b\dagger}(\mathbf{i})h^a(\mathbf{i})\big)$ are $SU(4)$ generators under $\mathbf{6}$ irrep, $\mathbf{r}_{\left<\mathbf{i}\mathbf{j}\right>}$ denotes the adjacent sites of bond $\left<\mathbf{ij}\right>$, and the Cooper pair density satisfies $n_h(\mathbf{i})=1-n_q(\mathbf{i})$. In \Eq{eq:h}, since the nearest-neighbor antiferromagnetic Heisenberg interaction favors two-site $SU(4)$ singlets, upon decreasing $\epsilon$, the singlet pair-superfluid state of $h^a(\mathbf{i})$~\footnote{In addition, the attractive nearest-neighbor density-density interaction also favors pair-superfluid states}, which is exactly the singlet charge-$4e$ SC state of electrons, can be proliferated against the $h^a(\mathbf{i})$ condensed state~\cite{zhang2025psf}. Physically, at low doping, the motion of Cooper pairs $h^a(\mathbf{i})$ is strictly constrained as they are local excitations. On the other hand, the quartets are nonlocal, suggesting they be efficiently condensed, similar to the mechanism of kinetic energy driven SC in bilayer nickelates~\cite{ESDreview}. This implies that, without net attraction ($\epsilon\ge 0$), doping the featureless Mott insulator with $SU(4)$ symmetry formed at $\nu=4$ should give rise to a primary singlet charge-$4e$ SC phase. In the next section we perform density matrix renormalization group (DMRG)~\cite{Hauschild2024} simulations to verify these intuitions for both $SU(4)$ and $Sp(4)$ ESD models.

\begin{figure}[h]
    \centering
\includegraphics[width=1.\linewidth]{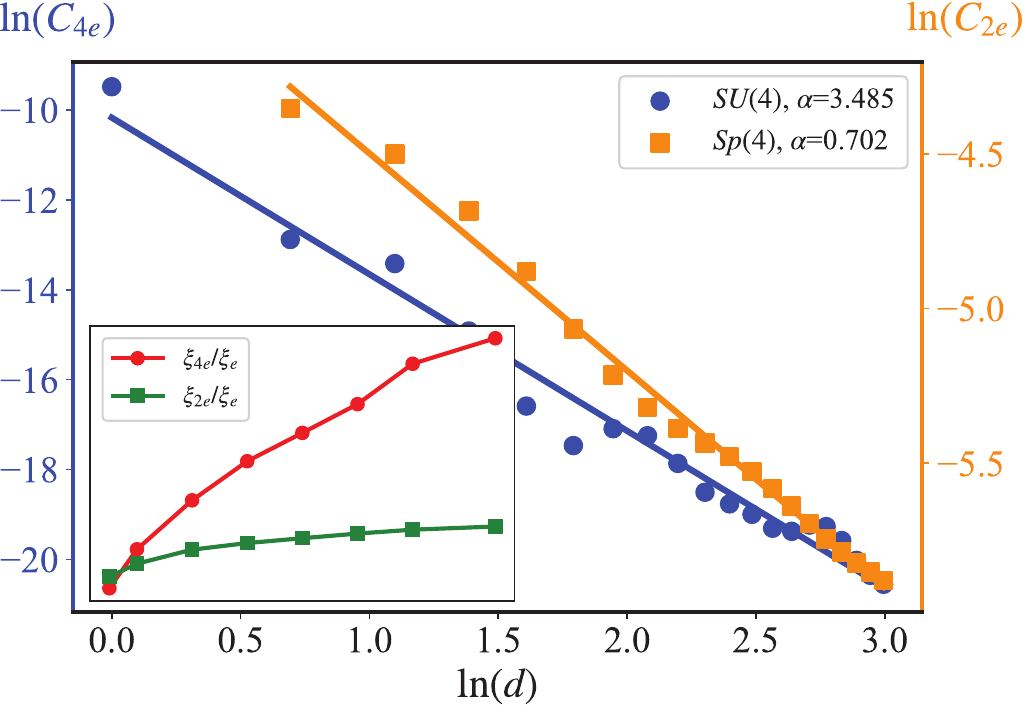}
\caption{Infinite DMRG simulation results for ESD model \Eq{eq:bH} with $L_x=8$. Here $x = 0.25$ and $\epsilon=0$  with $r_1=r_5=2/\sqrt{3}$ for $SU(4)$ ESD model, and for $Sp(4)$ model $x=0.125$, $\epsilon=0$, $\delta=0.4t$, and $r_1=0.828$, $r_5=1.15$ corresponding to $J_V/J_L=0.95$. The plot shows the log-log plot of the quartet-quartet correlation function $C_{4e}(d)$ for $SU(4)$ (blue) and pair-pair correlation function $C_{2e}(d)$ for $Sp(4)$ (orange) against distance $d$. The bond dimension is $m=15000$ for $SU(4)$ model and $m=8000$ for $Sp(4)$ model. Inset: correlation length $\xi/\xi_e$ with respect to different bond dimensions for $SU(4)$ ESD model, where $\xi_{4e(2e)}$ corresponds to the correlation length of quartetting (pairing), and $\xi_e$ is the single particle correlation length.
}
\label{fig:SU4_ESD}
\end{figure}

\textit{Superconducting phase: $SU(4)$ versus $Sp(4)$.---} In the infinite DMRG simulation for \Eq{eq:esd} on a one-dimensional chain, we consider the case with no net attractive interaction ($\epsilon=0$). The doping measured from half-filling ($\nu=4$) is captured by $x$, where $x= 2-\nu/2$.  The simulation results on $SU(4)$ ESD model at $x=0.25$ are shown in Fig.~[\ref{fig:SU4_ESD}]. We find that, the correlation length of the singlet quartetting order parameter grows fast as the increasing of the bond dimension, and always dominates that of the vector pairing. In addition, the correlation function of the quartetting order parameter has a power law decay, while that of the pairing decays exponentially. These confirm the onset of a primary charge-$4e$ SC phase. Upon increasing the doping $x$, the charge-$4e$ SC gets suppressed by the onsite Hubbard repulsion, and the flavor-polarized charge-$2e$ SC phase emerges~\cite{supp}.

\begin{figure}[h]
    \centering
\includegraphics[width=0.83\linewidth]{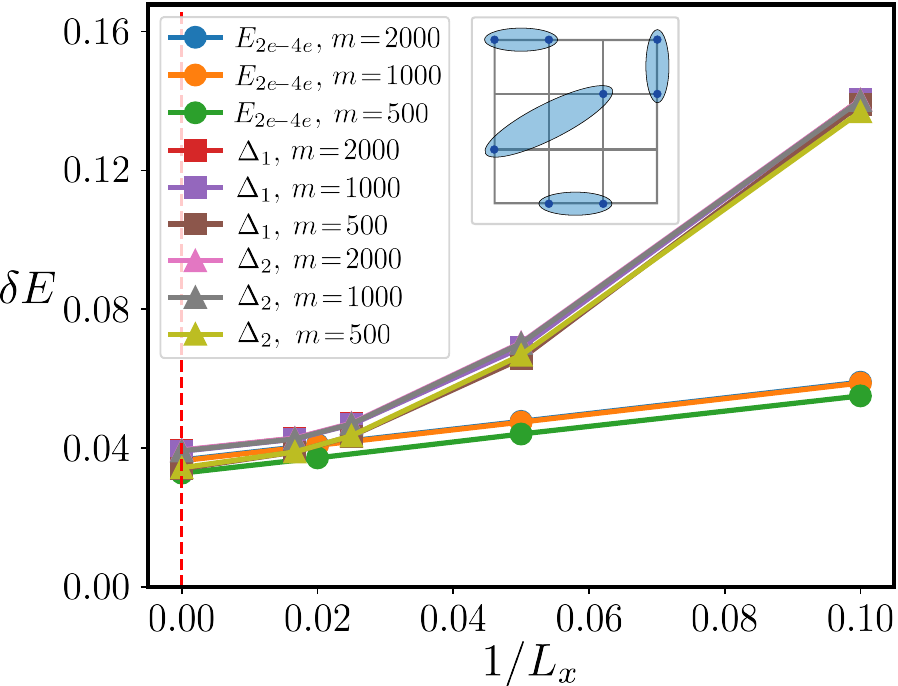}
\caption{ Finite DMRG simulation results of quartet binding energy and flavor gap in $SU(4)$ ESD model with $r_1=r_5=2/\sqrt{3}$ and $\epsilon=\delta=0$. Starting from the featureless Mott insulator at $x = 0$ ($\nu = 4$), we dope four holes into the system and compute the binding energy and flavor gap for fixed $L_y = 10$ and $L_x = 10, 20, 40, 50, 60$. The results are well converged with respect to the bond dimension. The extrapolation to $L_x\rightarrow+\infty$ shows that both the binding energy and the flavor gap remain finite in the thermodynamic limit along the $x$ direction, suggesting the formation of a stable quartet and the existence of a primary charge-$4e$ SC phase. The inset shows the nonlocal nature of the quartets, where the onsite Cooper pairs are marked as blue dotes, and quartets are circled by light blue ellipses.
}
\label{fig:SU4_ESD_finite_DMRG}
\end{figure}

To further verify the existence of the primary charge-$4e$ SC phase in the $SU(4)$ ESD model in the two-dimensional limit, we perform finite DMRG simulations for the quartet binding energy $E_{2e-4e}$ and the flavor gap $\Delta$ by starting from the featureless Mott insulator at $x = 0$ ($\nu = 4$) and doping four holes into the system. The results are shown in Fig.~[\ref{fig:SU4_ESD_finite_DMRG}]. Via extrapolation to thermodynamic limit, we find both the binding energy and the flavor gap are finite and positive. These suggest stable quartet formation upon doping and support the primary charge-$4e$ SC phase. In addition, we also find a relatively small charge-$2e$ gap, implying possible pseudogap metal phase when temperature exceeds this gap.

On the other hand, for the $Sp(4)$ ESD model, the infinite DMRG simulation is performed at doping $x=0.125$ on a one-dimensional chain, with $r_1=0.828$ and $r_5=1.15$ corresponding to $J_V/J_L=0.95$. Here, the correlation function of Cooper pairs exhibits power law decay, as shown in Fig.~[\ref{fig:SU4_ESD}], which indicates a primary charge-$2e$ SC. This $Sp(4)$ singlet charge-$2e$ SC phase exists in $0\le x\le 1$~\cite{supp}, similar to that in the $SU(2)$ ESD model~\cite{ESDreview}.

\textit{Normal state and finite temperature phase diagram.---} The SC phases in $SU(4)$ and $Sp(4)$ ESD models differ drastically. This naturally leads to the inquiry of the physical property of their corresponding normal states at finite temperature $T$. In the $SU(2)$ ESD model, it is shown that the normal state can be a conventional Fermi liquid corresponding to the free fermion, or a ``second Fermi liquid" whose Fermi surface cannot be adiabatically connected to the conventional one~\cite{ESDreview}, both of which have Fermi surface volume allowed by a flux insertion argument~\cite{Oshikawa2000}. This flux insertion argument can be also applied to the $SU(4)$ and $Sp(4)$ ESD models, where symmetries limit the Fermi surface volume $\mathcal{A}$. 
First, the charge $U(1)$ symmetry in each layer yields $ \sum_{\sigma}\mathcal{A}_{l,\sigma}=\nu$ (mod $1$). The $SU(4)$ or $Sp(4)$ flavor rotation symmetry further requires $\mathcal{A}_{l,\sigma}=\mathcal{A}_{l,\sigma^\prime}$ due to Schur's lemma, while the layer exchange symmetry holds $\mathcal{A}_{+,\sigma}=\mathcal{A}_{-,\sigma}$. Thus for all $l,\sigma$ the Fermi surfaces have equal volume $\mathcal{A}$, leading to four consistent solutions $\mathcal{A}=\nu/8-\upsilon/4$ with $\upsilon=0,1,2,3$. The corresponding Fermi liquid phases are denoted as FL$\upsilon$, all of which can be realized in $SU(4)$ and $Sp(4)$ ESD models. When $\nu \approx 2\upsilon$, the interlayer ordering is involved with $2\upsilon$ electrons. Thus, the electron density is $\rho_e=\nu-2\upsilon$, which exactly produces $\mathcal{A}=\nu/8-\upsilon/4$. In particular, FL0 is corresponding to the free fermion, which should appear at $\nu\gtrsim 0$ ($x\lesssim 2$). Other three Fermi liquids are not adiabatically connected to the free fermion case, similar to that in the $SU(2)$ ESD model. In particular, the primary charge-$4e$ SC phase should be viewed as a descendant of the intrinsically strongly correlated Fermi liquid FL2.

\begin{figure}[h]
    \centering
\includegraphics[width=1.\linewidth]{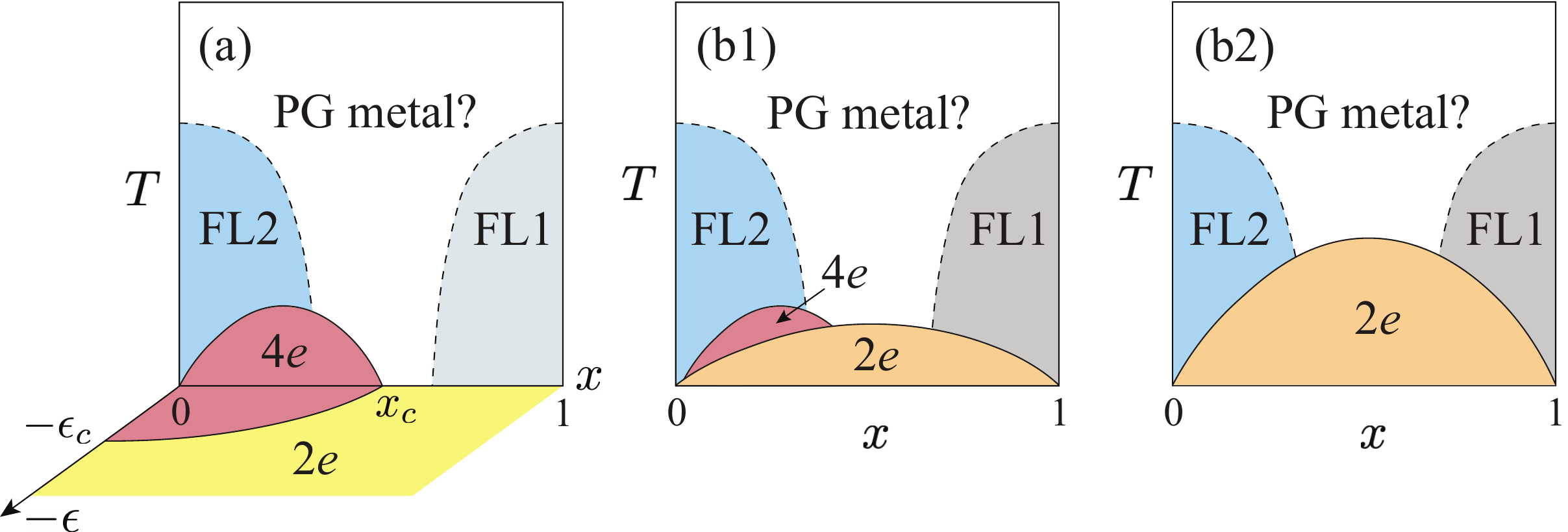}
\caption{Schematic phase diagram of (a) $SU(4)$ ESD model ($\delta= 0$), and (b) $Sp(4)$ ESD model with (b1) $\delta\gtrsim 0$, and (b2) $\delta\gg t,|\epsilon |$. In (b) the $\epsilon$-axis is neglected since the zero temperature phase is always a singlet charge-$2e$ SC.}
\label{fig:pd}
\end{figure}

The phase diagram as a function of doping $x$, temperature $T$, and boson energy $\epsilon$ for $SU(4)$ ESD model is presented in Fig.~[\ref{fig:pd}.(a)]. Here negative $\epsilon$ reflects net interlayer attraction contributed by negative $v$ in \Eq{eq:bH}. At zero temperature, the primary charge-$4e$ SC phase exists for $x<x_c\approx0.5$ and $\epsilon>\epsilon_c\approx0$ as revealed in the DMRG simulations in Figs.~[\ref{fig:SU4_ESD}] and [\ref{fig:SU4_ESD_finite_DMRG}]~\footnote{In the ESD model, the featureless Mott insulator appears around $x\gtrsim0$, which should occupy a very small region and not shown explicitly on the phase diagram.}. For $x>x_c$ or $\epsilon<\epsilon_c$, the flavor-polarized charge-$2e$ SC phase appears. As it spontaneously breaks the $SU(4)$ symmetry to $Sp(4)$~\footnote{The five Goldstone modes are due to flavor symmetry breaking and not corresponding to the charge $U(1)$ symmetry. Thus they are not absorbed by the electromagnetic fields, and remain gapless in the system.}, it cannot extend to finite temperature according to the Mermin--Wagner theorem~\cite{MW}. However, at finite temperature, formation of local Cooper pairs may lead to a pseudogap (PG) metal phase. Fermi liquid phases FL1 and FL2 should occur around $x=1$ and $x=0$, respectively, as parent phases of the charge-$2e$ and the charge-$4e$ SC phases~\footnote{FL1 should be understood as a FL$^*$ phase as it contains dynamical gauge field.}.

The phase diagrams for $Sp(4)$ ESD model are shown in Fig.~[\ref{fig:pd}.(b)]. The $\epsilon$-axis is neglected since the zero temperature phase is always a singlet charge-$2e$ SC, as seen from the infinite DMRG results~\cite{supp}~\footnote{Due to the restriction of computational resources, we do not exclude the possibility that charge-$4e$ SC phase may still persist for sufficiently small $x$.}. For finite temperature, we still expect charge-$4e$ SC to exist when $\delta\gtrsim 0$ and the $SU(4)$ symmetry is nearly restored. As a result, the phase diagram should be similar to that of the $SU(4)$ ESD model, as shown in Fig.~[\ref{fig:pd}.(b1)]. It is crucial to note that, with $Sp(4)$ symmetry, the charge-$4e$ SC phase only occurring at finite temperature becomes a vestigial order of the singlet charge-$2e$ SC phase. As $\delta$ is increased, the charge-$4e$ region shrinks and disappears at some critical $\delta_c$, leaving FL2 and FL1 phases at $x\gtrsim 0$ and $x\lesssim 1$, respectively.  The phase diagram in the limit $\delta\gg t,\epsilon$ is shown in Fig.~[\ref{fig:pd}.(b2)]. It shows a similar behavior for $x\gtrsim 0$ and $x\lesssim 1$, resembles the phase diagram of the $SU(2)$ ESD model. These can be understood as follows. As $\delta\gg t,\epsilon$, the quintet states $\ket{\gamma^a(\mathbf{i})}$ effectively decouples from the system, leaving two singlets in low energy only. Thus the bosonic part of the model recovers that in the $SU(2)$ ESD model~\cite{ESDreview}, yielding a similar phase diagram. In addition, in this limit, upon exchanging the two singlets the Hamiltonian \Eq{eq:esd} and the projected electron operators \Eq{eq:ele} remain qualitatively unchanged, suggesting the similarity between $x\gtrsim 0$ and $x\lesssim 1$.

Phase transitions between charge-$4e$ and charge-$2e$ SC phases differ in the $SU(4)$ and $Sp(4)$ ESD models. For $SU(4)$ case, the condensation of Cooper pairs spontaneously breaks both the $SU(4)$ flavor symmetry to $Sp(4)$, and the $\mathbb{Z}_4$ charge conservation symmetry to $\mathbb{Z}_2$, where the latter one is an Ising transition~\cite{Wu2005,Gao2022,supp}. For $Sp(4)$ case, the order parameter of the singlet charge-$2e$ SC is a component of the above flavor-polarized charge-$2e$ SC, as the singlet state $\ket{\eta(\mathbf{i})}$ is a component of the $\mathbf{6}$ vector irrep of $SU(4)$, and its transition to the charge-$4e$ SC phase is also an Ising one inherited from the $SU(4)$ case.

\textit{Conclusions and outlooks.---} In this work we construct a concrete lattice model to show primary charge-$4e$ superconducting phase at zero temperature from doping an $SU(4)$ symmetric featureless Mott insulator. Our results highlight that neither enlarged internal symmetry nor strong correlations alone are sufficient to stabilize a primary charge-$4e$ superconductivity. It is the combined effect of (i) kinematic constraints from the $SU(4)$ center forbidding the singlet charge-$2e$ pairing channel, and (ii) dynamical constraints from Mott physics and restricted particle motions in the ESD model, that renders primary charge-$4e$ superconductivity to emerge. These can be straightforwardly generalized to $SU(2n)$ or $Sp(2n)$ cases. In the $SU(2n)$ ESD model, we expect a primary singlet charge-$2ne$ SC phase arising from slightly doping the featureless Mott insulator at half-filling $\nu = 2n$, while sufficiently large doping is expected to favor various symmetry-breaking charge-$2e$ SC phases. In contrast, in the $Sp(2n)$ ESD model, a singlet charge-$2e$ superconductor should remain the primary superconducting phase.

From an experimental perspective, the parent bilayer Hubbard models with onsite $SU(2n)$ or $Sp(2n)$ symmetry may be engineered in several multicomponent fermion platforms. Ultracold alkali and alkaline earth atoms in optical lattices naturally realize high symmetries~\cite{Cazalilla2009,Gorshkov2010,killian2010,wu2010,wu2012,Taie2012}, and bilayer geometries with controllable interlayer couplings could implement the ESD-type constructions discussed here. In solid state systems, approximate $SU(4)$ or $Sp(4)$ internal symmetries can emerge from spin, valley, and layer degrees of freedom in moir\'e heterostructures~\cite{You2019,Bernevig2021,Wang2021topo,Khalaf2022,Chichinadze2022,Onari2022,Parthenios2023}, providing candidate settings for primary charge-$4e$ superconductivity.

\textit{Acknowledgements.---} We acknowledge Fa Wang, Eslam Khalaf, Zhehao Dai, Zhaoyu Han, Pavel Nosov, Clemens Kuhlenkamp, and Hanbit Oh for helpful discussions. ZQG acknowledges support from Berkeley graduate program. YQW is supported by the JQI postdoctoral fellowship at the University of Maryland. YHZ is supported by the Alfred P. Sloan Foundation through a Sloan Research Fellowship.

\bibliography{4e.bib}

\clearpage

\onecolumngrid

\vspace{0.3cm}

\supplementarysection
 
\begin{center}
\Large{\bf Supplemental Material for ``Primary charge-$4e$ superconductivity from doping a featureless Mott insulator"}
\end{center}

\section{Proof of the center enforcement mechanism}

For the simply connected classical Lie group $G$, its center, denoted as $Z(G)$, has
\beq
Z(SU(N))=\mathbb{Z}_N,\quad Z(Sp(2N))=\mathbb{Z}_2,\quad Z(\mathrm{Spin}(N))=\left\{\begin{array}{rl}
   \mathbb{Z}_2,  &  N\text{ odd}\\
   \mathbb{Z}_2\times \mathbb{Z}_2,  & N=0\text{ mod }4\\
   \mathbb{Z}_4,  & N=2\text{ mod }4\\
\end{array}\right. .
\eeq
Restricted to the center, the fundamental spinor irrep $\Psi$ becomes a character $\chi$ carrying unit charge under the center, as summarized:
\beq
\chi_{SU(N)}=e^{\frac{2\pi i}{N}},\quad\chi_{Sp(2N)}=-1,\quad
\chi_{\mathrm{Spin}(N)}=\left\{\begin{array}{rl}
   -1,  &  N\text{ odd}\\
   (-1,-1),  & N=0\text{ mod }4\\
   i,  & N=2\text{ mod }4\\
\end{array}\right. .
\eeq
Thus, $\chi^N=1$ for $SU(N)$, $\chi^4=1$ for $\mathrm{Spin}(4N\!+\!2)$ with $N=2$ mod $4$, and $\chi^2=1$ for other groups. Among the exceptional Lie groups, only $E_6$ and $E_7$ have nontrivial centers $Z(E_6)=\mathbb{Z}_3$ and $Z(E_7)=\mathbb{Z}_2$, while $Z(G_2)=Z(F_4)=Z(E_8)=\mathbb{Z}_1$. From the group theory perspective, as electrons typically transform under the fundamental (spinor) irrep, a singlet charge-$2ne$ SC state arises when the direct product of $2n$ such fundamental spinors contains singlets, {\it i.e.}, the trivial irrep, in the decomposition. Therefore, the most promising platform for singlet SC phases with arbitary large charge $Q$ is $SU(N)$ symmetric systems, as the $\mathbb{Z}_N$ center of $SU(N)$ enforces $2n=0$ mod $N$. This enforcement yields the primary singlet SC phase to have charge-$2ne$ for $SU(2n)$, or, $SU(n)$ with odd $n$ since the SC order parameter must be bosonic. A similar case of $SU(n)$ with odd $n$ is $E_6$, where the $\mathbb{Z}_3$ center enforces $2n=0$ mod $3$, suggesting primary singlet charge-$6e$ SC. Among spin groups, $\mathrm{Spin}(4N\!+\!2)$ has $\mathbb{Z}_4$ center that enforces $2n=0$ mod $4$ that can yield primary singlet charge-$4e$ SC regardless of $N$. For other spin groups $\mathrm{Spin}(N)$ ($N\neq 2$ mod $4$) and symplectic series $Sp(2N)$, as well as $G_2$, $F_4$, $E_7$, and $E_8$, primary SC phase with higher charge is unlikely.

\section{The $Sp(4)$ and $SU(4)$ algebra and the Hilbert space}

In this section we briefly review the $SU(4)$ and $Sp(4)$ algebra in 4-flavor fermion systems. We adopt the convention for the four by four gamma matrices as
\beq
\Gamma^1 =\begin{pmatrix}
0 & -iI\\
iI & 0
\end{pmatrix},\quad
\Gamma^{2,3,4} =\begin{pmatrix}
\vec{\sigma} & 0\\
0 & -\vec{\sigma}
\end{pmatrix},\quad
\Gamma^5 =\begin{pmatrix}
0 & I\\
I & 0
\end{pmatrix},
\eeq
where $I$ and $\vec{\sigma}$ are two by two identity and Pauli matrices. These five gamma matrices form an $Sp(4)$ vector. The generators of $Sp(4)$ are defined as $\Gamma^{ab}=-(i/2)[\Gamma^a, \Gamma^b]$, and the charge conjugation operator is defined as $R=\Gamma^1\Gamma^3$. They together form the generator set of $SU(4)$, which defines the charge-neutral fermion bilinear operators $M_\pm^{ab}(\mathbf{i})$ with $1\le a, b\le 6$: (for clarity we omit the layer index and use $\psi(\mathbf{i})$ to refer to electrons in either layer) 
\beq
&&M^{ab}(\mathbf{i})=L^{ab}(\mathbf{i})=\frac{1}{2}\psi^\dagger(\mathbf{i}) \Gamma^{ab} \psi(\mathbf{i}),\quad 1\le a, b\le 5,\\
&&M^{a6}(\mathbf{i})=V^a(\mathbf{i})=\frac{1}{2}\psi^\dagger(\mathbf{i}) \Gamma^a \psi(\mathbf{i}),\quad 1\le a\le 5,
\eeq
satisfying the commutation rule
\beq
[M^{ab}(\mathbf{i}),M^{cd}(\mathbf{i})]=i(\delta^{ad}M^{bc}(\mathbf{i})+\delta^{bc}M^{ad}(\mathbf{i})-\delta^{ac}M^{bd}(\mathbf{i})-\delta^{bd}M^{ac}(\mathbf{i})), \quad 1\le a, b, c, d\le 6.
\eeq
Thus, $\{M^{ab}(\mathbf{i})\}$ and $\{L^{ab}(\mathbf{i})\}$ are indeed the generators of $SU(4)$ and $Sp(4)$, respectively. 

In the Mott limit, the structure of the local Hilbert space on each rung with integer fillings can be understood from the group theory perspective. At $\nu=4$, each site hosts six states in the balanced filling sector. Note that, imbalance filling sectors with different electron numbers in the two layers also exist; however, they have much higher energies at even integer $\nu$ due to the on-site Hubbard interaction, and are not considered in low energy. The six on-site states in each layer are transformed in the $\mathbf{6}$ vector irrep of $SU(4)$.
For $Sp(4)$, they are further decomposed to the $\mathbf{5}$ vector irrep 
\beq
\ket{\gamma^a_\pm(\mathbf{i})}=-\frac{i}{2}c_{\pm,\sigma}^\dagger(\mathbf{i})(R\Gamma^a)_{\sigma\sigma^\prime}c^\dagger_{\pm,\sigma^\prime}(\mathbf{i})\ket{0}
\eeq
and the singlet 
\beq
\ket{\eta_\pm(\mathbf{\mathbf{i}})}=\frac{1}{2}c_{\pm,\sigma}^\dagger(\mathbf{i})R_{\sigma\sigma^\prime}c^\dagger_{\pm,\sigma^\prime}(\mathbf{i})\ket{0}.
\eeq
Due to the interlayer Heisenberg coupling, the total 36 states on each rung are divided into $SU(4)$ irreps as $\mathbf{6}\otimes \mathbf{6}=\mathbf{35}\oplus\mathbf{1}$, and $Sp(4)$ irreps as $(\mathbf{5}\oplus\mathbf{1})\otimes(\mathbf{5}\oplus\mathbf{1})=(\mathbf{15}\oplus\mathbf{10}\oplus\mathbf{1})\oplus(\mathbf{5}\oplus\mathbf{5})\oplus\mathbf{1}$. In the $Sp(4)$ case there are two quartetting singlets. One of them has the same wavefunction as the unique $SU(4)$ quartetting singlet 
\beq
\ket{q(\mathbf{i})}=\frac{1}{\sqrt{6}}\epsilon_{\mu\nu\rho\lambda}c_{+,\mu}^\dagger(\mathbf{i})c_{+,\nu}^\dagger(\mathbf{i})c_{-,\lambda}^\dagger(\mathbf{i})c_{-,\rho}^\dagger(\mathbf{i})\ket{0},
\eeq
while the other one $\ket{\tilde{q}(\mathbf{i})}$ composed from two charge-$2e$ singlets, $\mathbf{1}=\mathbf{1}\otimes\mathbf{1}$, has a more complicated form. For generic $J_{L,V}$, the ground state is the superposition of the two quartetting singlets, as discussed in detail in the next section.

At quarter filling $\nu=2$, each site hosts four states transformed under the fundamental spinor irrep $\mathbf{4}$ of $SU(4)$ or $Sp(4)$. With the interlayer Heisenberg coupling, the total 16 states on each rung are divided into $\mathbf{4}\otimes \mathbf{4}=\mathbf{10}\oplus\mathbf{6}$ for $SU(4)$, where the ground states transform under the $\mathbf{6}$ vector irrep. For $Sp(4)$, as $\mathbf{4}\otimes \mathbf{4}=\mathbf{10}\oplus\mathbf{5}\oplus\mathbf{1}$, rung singlet is allowed, denoted as $\ket{\eta(\mathbf{i})}$. Together with the $\mathbf{5}$ vector (rung quintet) $\ket{\gamma^a(\mathbf{i})}$, they transform under the $\mathbf{6}$ vector irrep of $SU(4)$. Similar physics occurs at $\nu=6$ due to the particle-hole symmetry. The wavefunctions of rung singlet and quintet states under $Sp(4)$ in this sector are
\beq
\ket{\eta(\mathbf{i})}=\ket{\gamma^6(\mathbf{i})}=\frac{1}{2}c_{+,\sigma}^\dagger(\mathbf{i})R_{\sigma\sigma^\prime}c^\dagger_{-,\sigma^\prime}(\mathbf{i})\ket{0},\quad\ket{\gamma^a(\mathbf{i})}=-\frac{i}{2}c_{+,\sigma}^\dagger(\mathbf{i})(\Gamma^aR)_{\sigma\sigma^\prime}c^\dagger_{-,\sigma^\prime}(\mathbf{i})\ket{0}.
\eeq

At fillings with odd $\nu$, however, imbalance filling is inevitable. For example, at $\nu=3$, the low energy sector has one and two electrons in the two sites of a rung. The site with one electron always hosts the fundamental spinor irrep $\mathbf{4}$ of $Sp(4)$ or $SU(4)$, while that with two electrons forms $\mathbf{5}\oplus\mathbf{1}$ or $\mathbf{6}$ under $Sp(4)$ or $SU(4)$, respectively. Thus, the lowest energy states are transformed under $\mathbf{4}=\mathbf{4}\otimes\mathbf{1}$ for $Sp(4)$ and $\bar{\mathbf{4}}\subset\mathbf{4}\otimes\mathbf{6}$ for $SU(4)$, respectively, both of which are four-dimensional. The layer exchange symmetry suggests the other four states degenerate with them. Similarly, for $\nu=1$ or $\nu=7$, the on-site low energy sector also contains two copies of four-dimensional irreps related by the layer exchange symmetry for both $SU(4)$ and $Sp(4)$ cases.

\section{Low energy states in the generalized ESD model}

\begin{figure}[h]
    \centering
\includegraphics[width=0.6\linewidth]{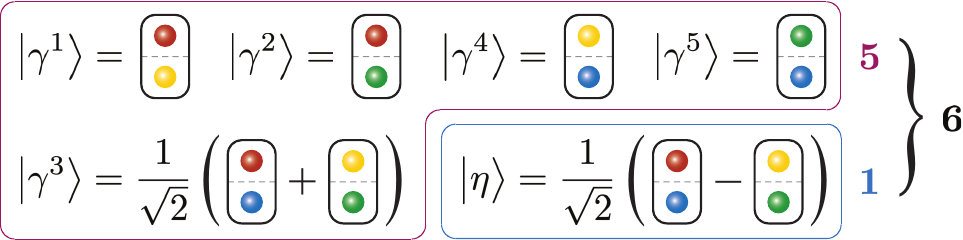}
\caption{States in $\nu=2$ sector organized into $SU(4)$ weight basis, where $\ket{\eta}$ naturally becomes the singlet under $Sp(4)$.}
\label{fig:statenew}
\end{figure}

The $SU(4)$ or $Sp(4)$ bilayer Hubbard model in the main text reads
\beq
H&=&-t\sum_{\left<\mathbf{i}\mathbf{j}\right>}c_{l,\sigma}^\dagger(\mathbf{i})c_{l,\sigma}(\mathbf{j})+\frac{u}{2}\sum_{\mathbf{i},l}\big(n_{l}(\mathbf{i})-2\big)^2+v\sum_{i}n_+({\bf i})n_-({\bf i})+J_L\sum_{\mathbf{i}}L^{ab}_+(\mathbf{i})L^{ab}_-(\mathbf{i})+J_V\sum_{\mathbf{i}}V^a_+(\mathbf{i})V^a_-(\mathbf{i}).
\eeq
In the Mott limit $u\gg J_{L,V}\gg t$ with doping from half-filling, the possible electron number per site per layer is 1 or 2. Thus, the total occupation number is $\nu=2,3,4$. With strong interlayer coupling $J_{L,V}$, we only keep the states with lowest energy in each sector, as discussed in the main text. For clarity we omit the site index $\mathbf{i}$ in wavefunctions.
\begin{itemize}
    \item Quartetting states $\ket{q}$:  $\nu=4$, $\epsilon_{q}=-4J_L(2+\sqrt{4 + 5(J_V/J_L)^2})+v$.
    \item Fermion states $\ket{f_{l,\sigma}}$: $\nu=3$, $\epsilon_f=-2J_L(2 +\sqrt{4 + 5 (J_V/J_L)^2})+u+2v$
    \item Pairing states $\ket{\eta}$ and $\ket{\gamma^a}$: $\nu=2$, $\epsilon_\eta=-5J_L(2-(J_V/J_L))+2u+4v$ and $\epsilon_{\gamma}=-2J_L(1+(3/2)(J_V/J_L))+2u+4v$.
\end{itemize}
In the $Sp(4)$ ESD model the $n=2$ state is the rung singlet, while in the $SU(4)$ ESD model, there are six degenerate states, $\ket{\eta}$ and $\ket{\gamma^a}$. The broken of $SU(4)$ is tuned by the energy difference $\delta>0$. The detailed wave functions are listed as follows.
In the $Sp(4)$ ESD model, the lowest energy $\nu=4$ state is an $Sp(4)$ singlet, with ($r=J_V/J_L$, unnormalized)
\bea
\ket{q}=&\Big(c^\dagger_{-,1}c^\dagger_{-,2}c^\dagger_{+,3}c^\dagger_{+,4}-c^\dagger_{-,1}c^\dagger_{-,3}c^\dagger_{+,2}c^\dagger_{+,4}-c^\dagger_{-,2}c^\dagger_{-,4}c^\dagger_{+,1}c^\dagger_{+,3}+c^\dagger_{-,3}c^\dagger_{-,4}c^\dagger_{+,1}c^\dagger_{+,2}\\
&+\frac{2+r-\sqrt{4+5r^2}}{2r}(c^\dagger_{-,1}c^\dagger_{-,4}c^\dagger_{+,1}c^\dagger_{+,4}+c^\dagger_{-,2}c^\dagger_{-,3}c^\dagger_{+,2}c^\dagger_{+,3})\\
&+\frac{-2+r+\sqrt{4+5r^2}}{2r}(c^\dagger_{-,1}c^\dagger_{-,4}c^\dagger_{+,2}c^\dagger_{+,3}+c^\dagger_{-,2}c^\dagger_{-,3}c^\dagger_{+,1}c^\dagger_{+,4})\Big)\ket{0}.
\eea
For $\nu=2$, there are one singlet and five vectors, which can be organized into $SU(4)$ weight basis convenient for numerics as (see Fig.~\ref{fig:statenew} for illustration),
\bea
    \ket{\gamma^1}=&\frac{1}{\sqrt{2}}(-c^\dagger_{-,3}c^\dagger_{+,4}+c^\dagger_{-,4}c^\dagger_{+,3})\ket{0},\\
    \ket{\gamma^2}=&\frac{1}{\sqrt{2}}(-c^\dagger_{-,2}c^\dagger_{+,4}+c^\dagger_{-,4}c^\dagger_{+,2})\ket{0},\\
    \ket{\gamma^3}=&\frac{1}{2}(-c^\dagger_{-,1}c^\dagger_{+,4}-c^\dagger_{-,2}c^\dagger_{+,3}+c^\dagger_{-,3}c^\dagger_{+,2}+c^\dagger_{-,4}c^\dagger_{+,1})\ket{0},\\
    \ket{\gamma^4}=&\frac{1}{\sqrt{2}}(-c^\dagger_{-,1}c^\dagger_{+,3}+c^\dagger_{-,3}c^\dagger_{+,1})\ket{0},\\
    \ket{\gamma^5}=&\frac{1}{\sqrt{2}}(-c^\dagger_{-,1}c^\dagger_{+,2}+c^\dagger_{-,2}c^\dagger_{+,1})\ket{0}.\\
    \ket{\gamma^6}=&\ket{\eta}=\frac{1}{2}(-c^\dagger_{-,1}c^\dagger_{+,4}+c^\dagger_{-,2}c^\dagger_{+,3}-c^\dagger_{-,3}c^\dagger_{+,2}+c^\dagger_{-,4}c^\dagger_{+,1})\ket{0}.
\eea
For $\nu=3$, there are totally eight states, labeled by the layer and the $Sp(4)$ flavor indices (unnormalized),
\bea
\ket{f_{+,1}}=&\Big(c^\dagger_{-,2}c^\dagger_{+,3}c^\dagger_{+,4}-c^\dagger_{-,3}c^\dagger_{+,2}c^\dagger_{+,4}+\frac{r+2-\sqrt{4+5r^2}}{2r}c^\dagger_{-,4}c^\dagger_{+,1}c^\dagger_{+,4}+\frac{r-2+\sqrt{4+5r^2}}{2r}c^\dagger_{-,4}c^\dagger_{+,2}c^\dagger_{+,3}\Big)\ket{0},\\
\ket{f_{+,2}}=&\Big(c^\dagger_{-,1}c^\dagger_{+,3}c^\dagger_{+,4}-c^\dagger_{-,4}c^\dagger_{+,1}c^\dagger_{+,3}-\frac{r-2+\sqrt{4+5r^2}}{2r}c^\dagger_{-,3}c^\dagger_{+,1}c^\dagger_{+,4}+\frac{r+2-\sqrt{4+5r^2}}{2r}c^\dagger_{-,3}c^\dagger_{+,2}c^\dagger_{+,3}\Big)\ket{0},\\
\ket{f_{+,3}}=&\Big(c^\dagger_{-,1}c^\dagger_{+,2}c^\dagger_{+,4}+c^\dagger_{-,4}c^\dagger_{+,1}c^\dagger_{+,2}-\frac{r-2+\sqrt{4+5r^2}}{2r}c^\dagger_{-,2}c^\dagger_{+,1}c^\dagger_{+,4}-\frac{r-2+\sqrt{4+5r^2}}{2r}c^\dagger_{-,2}c^\dagger_{+,2}c^\dagger_{+,3}\Big)\ket{0},\\
\ket{f_{+,4}}=&\Big(c^\dagger_{-,1}c^\dagger_{+,2}c^\dagger_{+,4}+\frac{-3r+\sqrt{4+5r^2}}{2(1+r)}c^\dagger_{-,1}c^\dagger_{+,1}c^\dagger_{+,4}-\frac{2-r+\sqrt{4+5r^2}}{2(1+r)}\big(c^\dagger_{-,2}c^\dagger_{+,1}c^\dagger_{+,3}-c^\dagger_{-,3}c^\dagger_{+,1}c^\dagger_{+,2}\big)\Big)\ket{0},\\
\ket{f_{-,1}}=&\Big(c^\dagger_{-,2}c^\dagger_{-,3}c^\dagger_{+,4}+\frac{-3r+\sqrt{4+5r^2}}{2(1+r)}c^\dagger_{-,1}c^\dagger_{-,4}c^\dagger_{+,4}-\frac{2-r+\sqrt{4+5r^2}}{2(1+r)}\big(c^\dagger_{-,2}c^\dagger_{-,4}c^\dagger_{+,3}-c^\dagger_{-,3}c^\dagger_{-,4}c^\dagger_{+,2}\big)\Big)\ket{0},\\
\ket{f_{-,2}}=&\Big(c^\dagger_{-,1}c^\dagger_{-,3}c^\dagger_{+,4}+c^\dagger_{-,3}c^\dagger_{-,4}c^\dagger_{+,1}-\frac{2(1+r)}{2-r+\sqrt{4+5r^2}}c^\dagger_{-,1}c^\dagger_{-,4}c^\dagger_{+,3}-\frac{2+r-\sqrt{4+5r^2}}{2r}c^\dagger_{-,2}c^\dagger_{-,3}c^\dagger_{+,3}\Big)\ket{0},\\
\ket{f_{-,3}}=&\Big(c^\dagger_{-,1}c^\dagger_{-,2}c^\dagger_{+,4}+c^\dagger_{-,2}c^\dagger_{-,4}c^\dagger_{+,1}-\frac{2(1+r)}{2-r+\sqrt{4+5r^2}}c^\dagger_{-,1}c^\dagger_{-,4}c^\dagger_{+,2}-\frac{2+r-\sqrt{4+5r^2}}{2r}c^\dagger_{-,2}c^\dagger_{-,3}c^\dagger_{+,2}\Big)\ket{0},\\
\ket{f_{-,4}}=&\Big(c^\dagger_{-,1}c^\dagger_{-,2}c^\dagger_{+,3}-c^\dagger_{-,1}c^\dagger_{-,3}c^\dagger_{+,2}+\frac{2+r-\sqrt{4+5r^2}}{2r}c^\dagger_{-,1}c^\dagger_{-,4}c^\dagger_{+,1}+\frac{-2+r+\sqrt{4+5r^2}}{2r}c^\dagger_{-,2}c^\dagger_{-,3}c^\dagger_{+,1}\Big)\ket{0}.
\eea
The $Sp(4)$ electron operators are defined from matrix elements $\bra{i}c_{l,\sigma}\ket{j}$, where $i,j$ are the states in the $Sp(4)$ ESD model.

The $SU(4)$ limit corresponds to $r=J_V/J_L=1$. For $\nu=4$, the only quartetting singlet is $\ket{q}=1/(\sqrt{6})\,\epsilon_{\mu\nu\rho\lambda}c^\dagger_{-,\mu}c^\dagger_{-,\nu}c^\dagger_{+,\rho}c^\dagger_{+,\lambda}\ket{0}.$ For $\nu=2$, the six states become degenerate, $\{\ket{\eta}=\ket{\gamma^6},\ket{\gamma^1},\ket{\gamma^2},\ket{\gamma^3},\ket{\gamma^4},\ket{\gamma^5}\}$.
For $\nu=3$, the $8$ fermionic states are reduced to
\bea
    \ket{f_{+,1}}=&\frac{1}{\sqrt{3}}(c^\dagger_{-,2}c^\dagger_{+,3}c^\dagger_{+,4}-c^\dagger_{-,3}c^\dagger_{+,2}c^\dagger_{+,4}+c^\dagger_{-,4}c^\dagger_{+,2}c^\dagger_{+,3})\ket{0},\\
    \ket{f_{+,2}}=&\frac{1}{\sqrt{3}}(c^\dagger_{-,1}c^\dagger_{+,3}c^\dagger_{+,4}-c^\dagger_{-,3}c^\dagger_{+,1}c^\dagger_{+,4}+c^\dagger_{-,4}c^\dagger_{+,1}c^\dagger_{+,3})\ket{0},\\
    \ket{f_{+,3}}=&\frac{1}{\sqrt{3}}(c^\dagger_{-,1}c^\dagger_{+,2}c^\dagger_{+,4}-c^\dagger_{-,2}c^\dagger_{+,1}c^\dagger_{+,4}+c^\dagger_{-,4}c^\dagger_{+,1}c^\dagger_{+,2})\ket{0},\\
    \ket{f_{+,4}}=&\frac{1}{\sqrt{3}}(c^\dagger_{-,1}c^\dagger_{+,2}c^\dagger_{+,3}-c^\dagger_{-,2}c^\dagger_{+,1}c^\dagger_{+,3}+c^\dagger_{-,3}c^\dagger_{+,1}c^\dagger_{+,2})\ket{0},\\
    \ket{f_{-,1}}=&\frac{1}{\sqrt{3}}(c^\dagger_{-,2}c^\dagger_{-,3}c^\dagger_{+,4}-c^\dagger_{-,2}c^\dagger_{-,4}c^\dagger_{+,3}+c^\dagger_{-,3}c^\dagger_{-,4}c^\dagger_{+,2})\ket{0},\\
    \ket{f_{-,2}}=&\frac{1}{\sqrt{3}}(c^\dagger_{-,1}c^\dagger_{-,3}c^\dagger_{+,4}-c^\dagger_{-,1}c^\dagger_{-,4}c^\dagger_{+,3}+c^\dagger_{-,3}c^\dagger_{-,4}c^\dagger_{+,1})\ket{0},\\
    \ket{f_{-,3}}=&\frac{1}{\sqrt{3}}(c^\dagger_{-,1}c^\dagger_{-,2}c^\dagger_{+,4}-c^\dagger_{-,1}c^\dagger_{-,4}c^\dagger_{+,2}+c^\dagger_{-,2}c^\dagger_{-,4}c^\dagger_{+,1})\ket{0},\\
    \ket{f_{-,4}}=&\frac{1}{\sqrt{3}}(c^\dagger_{-,1}c^\dagger_{-,2}c^\dagger_{+,3}-c^\dagger_{-,1}c^\dagger_{-,3}c^\dagger_{+,2}+c^\dagger_{-,2}c^\dagger_{-,3}c^\dagger_{+,1})\ket{0}.
\eea
The $SU(4)$ electron operators are,
\bea
    c_{+,1}=&-\frac{\sqrt{2}}{\sqrt{3}}\ket{\gamma^1}\bra{f_{+,2}}-\frac{\sqrt{2}}{\sqrt{3}}\ket{\gamma^2}\bra{f_{+,3}}-\frac{\sqrt{2}}{\sqrt{3}}\ket{\gamma^4}\bra{f_{+,4}}+\frac{1}{\sqrt{2}}\ket{f_{-,1}}\bra{q},\\
    c_{+,2}=&-\frac{\sqrt{2}}{\sqrt{3}}\ket{\gamma^1}\bra{f_{+,1}}+\frac{\sqrt{2}}{\sqrt{3}}\ket{\gamma^3}\bra{f_{+,3}}+\frac{\sqrt{2}}{\sqrt{3}}\ket{\gamma^5}\bra{f_{+,4}}-\frac{1}{\sqrt{2}}\ket{f_{-,2}}\bra{q},\\
    c_{+,3}=&\frac{\sqrt{2}}{\sqrt{3}}\ket{\gamma^2}\bra{f_{+,1}}+\frac{\sqrt{2}}{\sqrt{3}}\ket{\gamma^3}\bra{f_{+,2}}-\frac{\sqrt{2}}{\sqrt{3}}\ket{\gamma^6}\bra{f_{+,4}}+\frac{1}{\sqrt{2}}\ket{f_{-,3}}\bra{q},\\
    c_{+,4}=&-\frac{\sqrt{2}}{\sqrt{3}}\ket{\gamma^4}\bra{f_{+,1}}-\frac{\sqrt{2}}{\sqrt{3}}\ket{\gamma^5}\bra{f_{+,2}}-\frac{\sqrt{2}}{\sqrt{3}}\ket{\gamma^6}\bra{f_{+,3}}-\frac{1}{\sqrt{2}}\ket{f_{-,4}}\bra{q},\\
    c_{-,1}=&-\frac{\sqrt{2}}{\sqrt{3}}\ket{\gamma^1}\bra{f_{-,2}}-\frac{\sqrt{2}}{\sqrt{3}}\ket{\gamma^2}\bra{f_{-,3}}-\frac{\sqrt{2}}{\sqrt{3}}\ket{\gamma^4}\bra{f_{-,4}}+\frac{1}{\sqrt{2}}\ket{f_{+,1}}\bra{q},\\
    c_{-,2}=&-\frac{\sqrt{2}}{\sqrt{3}}\ket{\gamma^6}\bra{f_{-,1}}+\frac{\sqrt{2}}{\sqrt{3}}\ket{\gamma^3}\bra{f_{-,3}}+\frac{\sqrt{2}}{\sqrt{3}}\ket{\gamma^5}\bra{f_{-,4}}-\frac{1}{\sqrt{2}}\ket{f_{+,2}}\bra{q},\\
    c_{-,3}=&\frac{\sqrt{2}}{\sqrt{3}}\ket{\gamma^2}\bra{f_{-,1}}+\frac{\sqrt{2}}{\sqrt{3}}\ket{\gamma^3}\bra{f_{-,2}}-\frac{\sqrt{2}}{\sqrt{3}}\ket{\gamma^6}\bra{f_{-,4}}+\frac{1}{\sqrt{2}}\ket{f_{+,3}}\bra{q},\\
    c_{-,4}=&-\frac{\sqrt{2}}{\sqrt{3}}\ket{\gamma^4}\bra{f_{-,1}}-\frac{\sqrt{2}}{\sqrt{3}}\ket{\gamma^5}\bra{f_{-,2}}-\frac{\sqrt{2}}{\sqrt{3]}}\ket{\gamma^6}\bra{f_{-,3}}-\frac{1}{\sqrt{2}}\ket{f_{+,4}}\bra{q}
\eea

After projecting the bilayer Hubbard model into the basis defined above, the effective Hamiltonian reads
\beq
    H_\mathrm{ESD}=-t\sum_{\langle \mathbf{i}\mathbf{j} \rangle}\mathcal{P}c^\dagger_{l,\sigma}(\mathbf{i})c_{l,\sigma}(\mathbf{i})\mathcal{P}+\sum_{i}\epsilon_q n_q(\mathbf{i})+\epsilon_\eta n_\eta(\mathbf{i})+\epsilon_\gamma n_\gamma(\mathbf{i})+\epsilon_f n_f(\mathbf{i}),
\eeq
with the constraints $n_q(\mathbf{i})+ n_\eta(\mathbf{i})+ n_\gamma(\mathbf{i})+n_f(\mathbf{i})=1$ and $2n_\eta(\mathbf{i})+2n_\gamma(\mathbf{i})+n_f(\mathbf{i})=2x$. As a result, there are only two tuning parameters, $\epsilon$ and $\delta$, in the generalized ESD model discussed in the main text
\beq
H_\mathrm{ESD}=-t\sum_{\left<\mathbf{i}\mathbf{j}\right>}\mathcal{P}c_{l,\sigma}^\dagger(\mathbf{i})c_{l,\sigma}(\mathbf{j})\mathcal{P}+\sum_\mathbf{i}\epsilon \big(n_q(\mathbf{i})+n_\eta(\mathbf{i})\big)+(\epsilon+\delta)n_\gamma(\mathbf{i}).
\eeq

\section{Cooper pair picture and field theory analysis}

\begin{figure}[h]
    \centering
\includegraphics[width=0.75\linewidth]{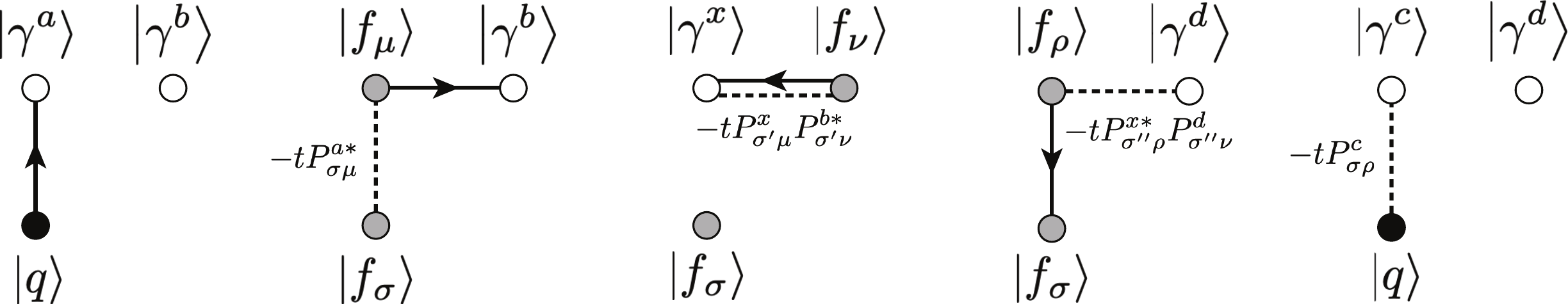}
\caption{Illustration for $q$-assisted exchange. Each arrowed solid line represents a step of virtual electron hopping, while its generating amplitude is marked by a dashed line. It can be directly read out that the total exchange amplitude from this process is $t^4\trace{P^{a\dagger}P^cP^{x\dagger}P^dP^{b\dagger}P^x}$.} 
\label{fig:hop}
\end{figure}

When $\epsilon\ll 0$, the generalized ESD model \Eq{eq:esd} can be reduced to an effective model of Cooper pairs from a perturbative expansion in $t/\epsilon$. For the $SU(4)$ ESD model, on each site the Hilbert space is $7=6+1$ dimensions formed by $\ket{q(\mathbf{i})}$ and $\ket{\gamma^a(\mathbf{i})}$, $a=1,\cdots,6$ with energy $-\epsilon$. The Cooper pair operators are defined as $h^a(\mathbf{i})=\ket{q(\mathbf{i})}\bra{\gamma^a(\mathbf{i})}$. Their hopping $\ket{\gamma^a(\mathbf{i})}\ket{q(\mathbf{j})}\rightarrow \ket{q(\mathbf{i})}\ket{\gamma^a(\mathbf{j})}$ is generated from the virtual process mediated by a second order electron hopping
\beq
\ket{\gamma^a(\mathbf{i})}\ket{q(\mathbf{j})}\rightarrow \ket{f_{l,\mu}(\mathbf{i})}\ket{f_{l^\prime,\nu}(\mathbf{j})}\rightarrow\ket{q(\mathbf{i})}\ket{\gamma^a(\mathbf{j})},
\eeq
where the intermediate states $\ket{f_{l,\mu}(\mathbf{i})}$ have energy $0$. The electron operator in the low energy Hilbert space is
\beq
\mathcal{P}c_{l,\sigma}(\mathbf{i})\mathcal{P}=\ket{\gamma^a(\mathbf{i})}P^a_{\sigma\mu}\bra{f_{l,\mu}(\mathbf{i})}+\ket{f_{-l,\sigma}(\mathbf{i})}\bra{q(\mathbf{i})},
\eeq
where $P^a_{\sigma\mu}$ is the CG coefficient from the decomposition of irrep $\bar{\mathbf{4}}\subset\mathbf{4}\times \mathbf{6}$ satisfying $\trace{P^{a\dagger}P^b}=\delta^{ab}$ due to the $SU(4)$ symmetry. The operation of electron hopping on $\ket{\gamma^a(\mathbf{i})}\ket{q(\mathbf{j})}$ reads
\beq
&&\mathcal{P}c^\dagger_{l,\sigma}(\mathbf{i})c_{l,\sigma}(\mathbf{j})\mathcal{P}\ket{\gamma^a(\mathbf{i})}\ket{q(\mathbf{j})}=\mathcal{P}c^\dagger_{l,\sigma}(\mathbf{i})\mathcal{P}\ket{\gamma^a(\mathbf{i})}\ket{f_{-l,\sigma}(\mathbf{j})}=P^{a*}_{\sigma\mu}\ket{f_{l,\mu}(\mathbf{i})}\ket{f_{-l,\sigma}(\mathbf{j})},\\
&&\mathcal{P}c^\dagger_{l^\prime,\sigma^\prime}(\mathbf{i})c_{l^\prime,\sigma^\prime}(\mathbf{j})\mathcal{P}\ket{f_{l,\mu}(\mathbf{i})}\ket{f_{-l,\sigma}(\mathbf{j})}=\mathcal{P}c^\dagger_{-l,\sigma^\prime}(\mathbf{i})\mathcal{P}P^b_{\sigma^\prime\sigma}\ket{f_{l,\mu}(\mathbf{i})}\ket{\gamma^b(\mathbf{j})}=P^b_{\mu\sigma}\ket{q(\mathbf{i})}\ket{\gamma^b(\mathbf{j})}.
\eeq
Thus, the Cooper pair hopping is described by
\beq
H_h^0=-\frac{t^2}{2|\epsilon |}P^{a*}_{\sigma\mu}P^a_{\mu\sigma}h^{a\dagger}(\mathbf{i})h^a(\mathbf{j})=-\frac{t^2}{2|\epsilon |}h^{a\dagger}(\mathbf{i})h^a(\mathbf{j}).
\eeq
Next we consider the interaction terms. At the leading order of $t/|\epsilon|$, the nearest-neighbor interaction is generated by a $q$-assisted exchange, $\ket{q(\mathbf{0})}\ket{\gamma^a(\mathbf{i})}\ket{\gamma^b(\mathbf{j})}\rightarrow \ket{q(\mathbf{0})}\ket{\gamma^c(\mathbf{i})}\ket{\gamma^d(\mathbf{j})}$, where $\mathbf{0}$ is adjacent to $\mathbf{i}$ or $\mathbf{j}$. Without losing of generality, we assume $\mathbf{0}$ is adjacent to $\mathbf{i}$, and consider the following process generating a nearest-neighbor antiferromagnetic Heisenberg interaction (see Fig.~\ref{fig:hop} for illustration): 
\beq
\ket{q(\mathbf{0})}\ket{\gamma^a(\mathbf{i})}\ket{\gamma^b(\mathbf{j})}\rightarrow \ket{f(\mathbf{0})}\ket{f(\mathbf{i})}\ket{\gamma^b(\mathbf{j})}
\rightarrow \ket{f(\mathbf{0})}\ket{\gamma^x(\mathbf{i})}\ket{f(\mathbf{j})}
\rightarrow \ket{f(\mathbf{0})}\ket{f(\mathbf{i})}\ket{\gamma^d(\mathbf{j})}\rightarrow \ket{q(\mathbf{0})}\ket{\gamma^c(\mathbf{i})}\ket{\gamma^d(\mathbf{j})},\nn\\
\eeq
where $x=1,\cdots,6$. The effective interaction generated from this process is
\beq
H^\mathrm{int}_h&=&-\frac{t^4}{8|\epsilon|^3}\trace{P^{a\dagger}P^cP^{x\dagger}P^dP^{b\dagger}P^x}\ket{q(\mathbf{0})}\ket{\gamma^a(\mathbf{i})}\ket{\gamma^b(\mathbf{j})}\bra{q(\mathbf{0})}\bra{\gamma^c(\mathbf{i})}\bra{\gamma^d(\mathbf{j})}\nn\\
&=&\frac{t^4}{64|\epsilon |^3}n_q(\mathbf{0})\big(S^{ab}(\mathbf{i})S^{ab}(\mathbf{j})-3n_h(\mathbf{i})n_h(\mathbf{j})\big),\label{eq:Heff}
\eeq
where $S^{ab}(\mathbf{i})=i\big(h^{a\dagger}(\mathbf{i})h^b(\mathbf{i})-h^{b\dagger}(\mathbf{i})h^a(\mathbf{i})\big)$ are $SU(4)$ generators under $\mathbf{6}$ irrep, and $n_h(\mathbf{i})=h^{a\dagger}(\mathbf{i})h^a(\mathbf{i})=n_\gamma(\mathbf{i})=1-n_q(\mathbf{i})$ is used. From \Eq{eq:Heff} it can be concluded that the two-site singlet has the lowest energy due to the antiferromagnetic Heisenberg-type interaction, which favors a singlet pair-superfluid state of $h^a(\mathbf{i})$, i.e. the singlet charge-$4e$ SC state of electrons, upon decreasing $|\epsilon|$. We note that, other interactions, including frustrated (non-nearest-neighbor) Heisenberg interactions and density/spin-assisted hoppings can be also generated by certain virtual processes, which are believed to be less relevant in the study of superconductivity.

The phase transition between the flavor-polarized charge-$2e$ SC and the singlet charge-$4e$ SC can be understood from a field theory perspective. A convenient representation to describe the vector Cooper pair and its coupling to the quartetting singlet is to use the complex antisymmetric tensor field $\phi_{ab}=-\phi_{ba}$ with $1\le a,b\le 4$. The full Ginzburg-Landau theory reads
\beq
\mathcal{L}=|D_{2A}\phi_{ab}|^2+s_\phi|\phi_{ab}|^2+v_\phi|\phi_{ab}|^4+v_\epsilon|\epsilon_{abcd}\phi_{ab}\phi_{cd}|^2+|D_{4A}\Phi|^2+s_\Phi |\Phi|^2+v_\Phi |\Phi|^4-\lambda (\Phi^\dagger \epsilon_{abcd}\phi_{ab}\phi_{cd}+h.c.),\nn\\
\eeq
where $\Phi$ is the quartetting field and $A$ is the electromagnetic field. $s$ and $v$ are phenomenological coupling parameters, and a positive $\lambda$ imposing the singlet quartetting channel from the pairing of two vector Cooper pairs. As $s_{\phi,\Phi}>0$, both boson fields are gapped, suggesting a fermi liquid phase. The onset of charge-$4e$ SC requires $s_\Phi$ evolves to negative before $s_\phi$, where $\Phi$ is condensed but $\phi$ remains gapped. Denote $\Phi=\rho e^{i\theta}$ with $\rho>0$ and $\theta\in [0,2\pi)$, and do the gauge transformation $\phi_{ab}\mapsto\phi_{ab}e^{i\theta/2}$. The Ginzberg-Landau Lagrangian becomes
\beq
\mathcal{L}=|D_{2A-\frac{1}{2}\partial_\mu\theta }\phi_{ab}|^2+s_\phi|\phi_{ab}|^2+v_\phi|\phi_{ab}|^4+v_\epsilon|\epsilon_{abcd}\phi_{ab}\phi_{cd}|^2-\lambda\rho ( \epsilon_{abcd}\phi_{ab}\phi_{cd}+h.c.)+\rho^2(\partial_\mu\theta-4A)^2,
\eeq
where $\theta$ described the Goldstone mode from the spontaneous breaking of the charge $U(1)$ symmetry, which is absorbed by the electromagnetic field $A$. From the last term it can be also seen that there is a residual $\mathbb{Z}_4$ gauge redundancy in $A$, suggesting a residual charge $\mathbb{Z}_4$ symmetry in the charge-$4e$ SC. Consequently, $\phi_{ab}$ is charged 2 (mod 4) under $\mathbb{Z}_4$. In what followings we will omit the $\mathbb{Z}_4$-valued $A$ for clarity. The flavor $SU(4)$ symmetry remains intact after the condensation of singlet $\Phi$. To see its further broken by the condensation of $\phi_{ab}$, we rewrite $\phi_{ab}$ into real and complex parts, $\phi_{ab}=\phi_{ab}^R+i\phi_{ab}^I$. Since $\lambda,\rho>0$, $\phi^R_{ab}$ will always be condensed before $\phi^I_{ab}$. The effective Lagrangian for $\phi^R$ reads
\beq
\mathcal{L}=\partial_\mu\phi^R_{ab}\partial_\mu\phi^R_{ab}+s_\phi\phi^R_{ab}\phi^R_{ab}-2\lambda\rho\epsilon_{abcd}\phi^R_{ab}\phi^R_{cd}+v_\phi(\phi^R_{ab}\phi^R_{ab})^2+v_\epsilon(\epsilon_{abcd}\phi^R_{ab}\phi^R_{cd})^2.
\eeq
Thus, for $\lambda\rho$ large enough, $\phi^R_{ab}$ is condensed. For example, when $s_\phi=v_\phi=v_\epsilon=1$, $\left<\phi^R_{ab}\right>=\pm\sqrt{(\lambda\rho-1)/15}$ for $\lambda\rho>1$. Thus, the $SU(4)$ symmetry is spontaneously broken to $Sp(4)$, leaving a Goldstone manifold $SU(4)/Sp(4)\sim S^5$. Moreover, as $\phi_{ab}$ also carries charge 2 mod 4 under the residual charge $\mathbb{Z}_4$ symmetry, its condensation also breaks $\mathbb{Z}_4$ to $\mathbb{Z}_2$, suggesting the Ising universality class of the transition. In together, for $G=SU(4)$, the transition between the flavor-polarized charge-$2e$ SC and the singlet charge-$4e$ SC is an Ising and an $SU(4)$ broken to $Sp(4)$ transition in charge and flavor sector, respectively. For $G=Sp(4)$, both the charge-$4e$ SC and the charge-$2e$ SC are singlets. Thus, the transition between them can be easily captured in a generalized XY model formalism, and shown to be in the Ising universality class.

\section{more detailed DMRG simulation results}
\begin{figure}[h]
    \centering
\includegraphics[width=0.95\linewidth]{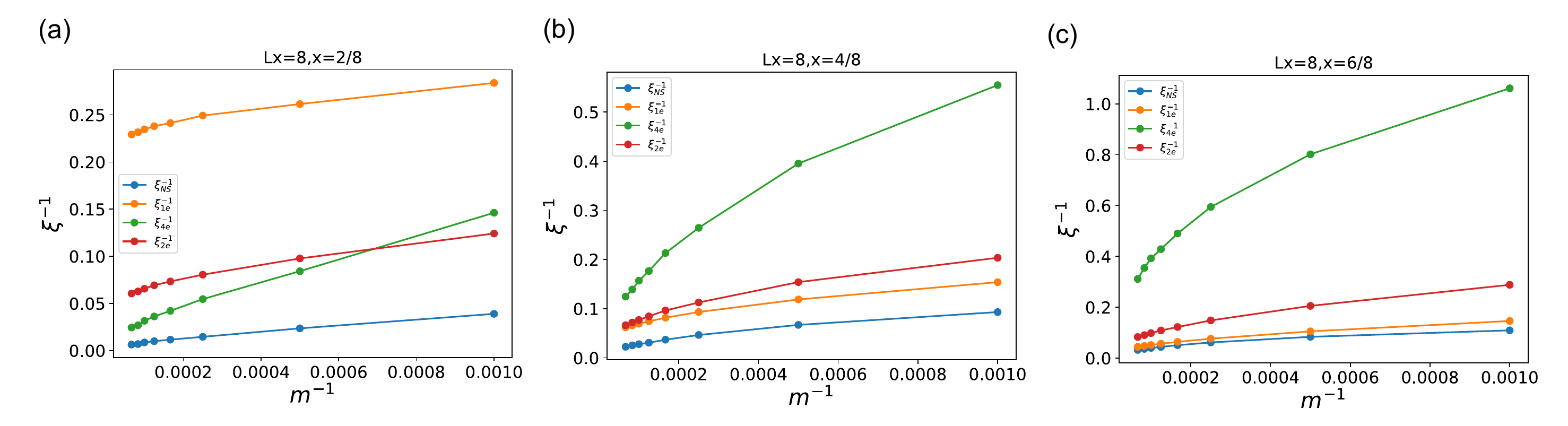}
\caption{Correlation length versus bond dimension in the $SU(4)$ ESD model for doping (a) $x=0.25$, (b) $x=0.5$ and (c) $x=0.75$, respectively.}
\label{fig:SU4_ESD_corr}
\end{figure}

In Fig.~\ref{fig:SU4_ESD_corr}, we show the correlation length in the $SU(4)$ ESD model for different dopings $x$For small doping $x=0.25$, the quartetting correlation length increases very fast as we increase the bond dimension, suggesting a charge-$4e$ SC phase.  For $x\ge 0.5$, the pairing correlation length dominates the quartetting correlation length, suggesting a flavor-polarized charge-$2e$ SC phase.

\begin{figure}[h]
    \centering
\includegraphics[width=0.66\linewidth]{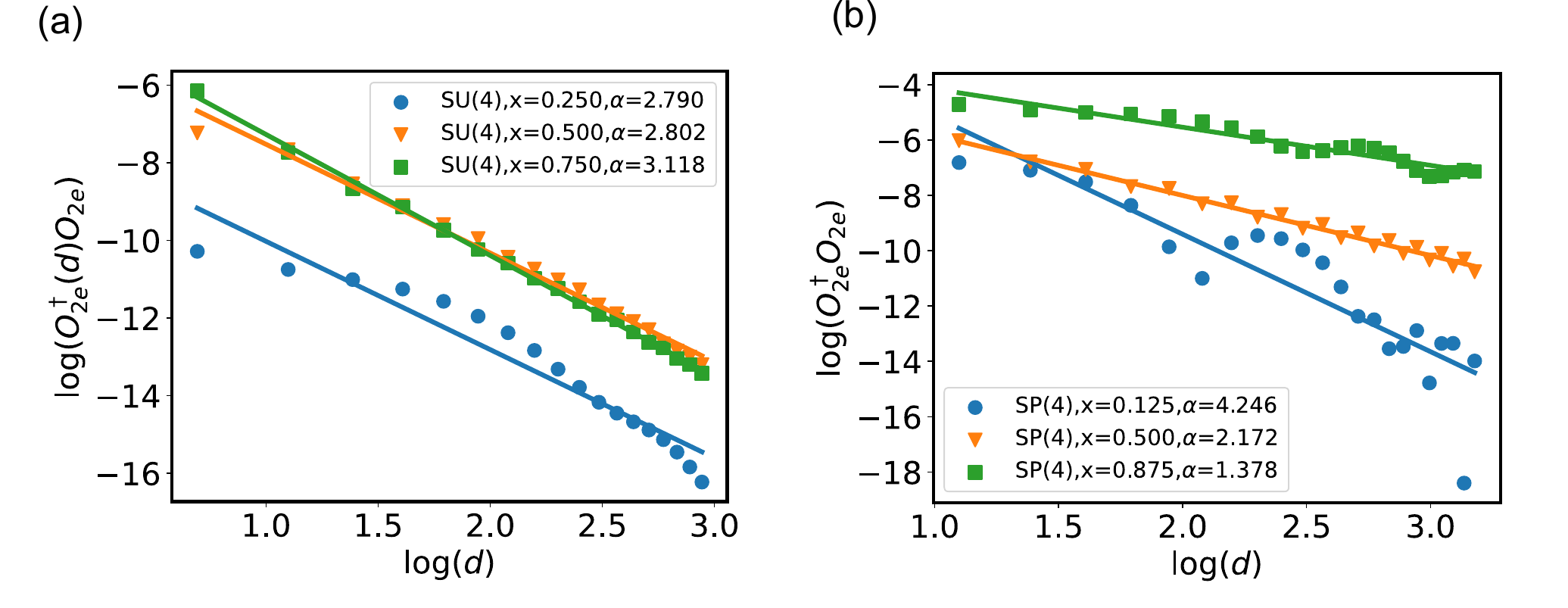}
\caption{The pair-pair correlation of (a) $SU(4)$ model, and (b) $Sp(4)$ model with $J_V/J_L=0$.}
\label{fig:2e_pair_pair}
\end{figure}

In Fig.~\ref{fig:2e_pair_pair}, we show the pair-pair correlation function of (a) $SU(4)$ ESD model, and (b) $Sp(4)$ ESD model with $\delta=\infty$ where the quintet states $\ket{\gamma^a(\mathbf{i})}$ is decoupled in the model. For $SU(4)$, The pair-pair correlation function has power law decay as $x\ge0.5$, which suggests a flavor-polarized charge-$2e$ SC phase, in consistent with the correlation length simulations.  For $Sp(4)$, the pair-pair correlation function always has power law decay as $x$ varies, suggesting a singlet charge-$2e$ SC phase regardless of $x$.

Data collected from finite DMRG simulations on $SU(4)$ ESD model are listed in Table~\ref{tab:finite_DMRG_single_electron} and \ref{tab:finite_DMRG_flavor}, showing the binding energies of quartet and Cooper pair, and the flavor gap, respectively.

\begin{table}[!htbp]
\begin{tabular}{cccc}
\hline\hline
$(L_x,L_y,\epsilon)$ & $E_{2e-1e}$ & $E_{4e-1e}$ & $E_{4e-2e}$\\
\hline
$(10,4,0)$&$-0.4689$&$-0.9641$&$-0.0263$\\
\hline
$(10,4,0.1)$&$-0.4225$&$-0.8666$&$-0.0216$\\
\hline
$(10,10,0)$&$-0.3516$&$-0.7621$&$-0.0590$\\
\hline
$(10,10,0.1)$&$-0.3106$&$-0.6784$&$-0.0571$\\
\hline
$(20,10,0)$&$-0.31669009$&$-0.681077$&$-0.04769682$\\
\hline
$(20,10,0.1)$&$-0.27635554$&$-0.59943687$&$-0.04672579$\\
\hline
$(50,10,0)$&$-0.30421911$&$-0.6494397723$&$-0.04100156079$\\
\hline
$(50,10,0.1)$&$-0.26395674$&$-0.56797891804$&$-0.0400654339890$\\
\hline\hline
\end{tabular}
\caption{Single-particle gap from finite DMRG simulation on $SU(4)$ ESD model. Here, $E_{2e-1e}=E(N_h=2)-E(N_h=0)-2(E(N_h=1)-E(N_h=0))$ with $N_h$ the number of doped holes, and similar definitions for $E_{4e-1e}$ and $E_{4e-2e}$. Via extrapolation to $L_x\rightarrow\infty$ we find $E_{4e-2e}= -0.036185$ for $\epsilon=0$ and $-0.035333$ for $\epsilon=0.1$. }\label{tab:finite_DMRG_single_electron}
\end{table}

\begin{table}[!htbp]
\begin{tabular}{ccc}
\hline\hline
$(L_x,L_y,\epsilon)$ & $\Delta_1$ & $\Delta_2$\\
\hline
$(10,4,0)$&$0.12375186$&$0.14816599$\\
\hline
$(10,4,0.1)$&$0.12111535$&$0.14727427$\\
\hline
$(10,10,0)$&$0.14044424$&$0.14038208$\\
\hline
$(10,10,0.1)$&$0.14315215$&$0.14308719$\\
\hline
$(20,10,0)$&$0.0691179$&$0.0703763$\\
\hline
$(20,10,0.1)$&$0.06831124$&$0.070244$\\
\hline
$(40,10,0)$&$0.04697752$&$0.04715717$\\
\hline
$(40,10,0.1)$&$0.04605805$&$0.04632073$\\
\hline
$(60,10,0)$&$0.04281439$&$0.04286323$\\
\hline
$(60,10,0.1)$&$0.0418815$&$0.04195449$\\
\hline\hline
\end{tabular}
\caption{Flavor gap from finite DMRG simulation on $SU(4)$ ESD model. Here, $\Delta_1$ and $\Delta_2$ correspond to flip one and two flavors, respectively. Extrapolation to $L_x\rightarrow\infty$, shows $\Delta_1=0.02817776$ and $\Delta_2=0.02751634$ for $\epsilon=0$, and $\Delta_1=0.02717070$ and $\Delta_2=0.02614442$ for $\epsilon=0.1$. }\label{tab:finite_DMRG_flavor}
\end{table}

\vfill 

\end{document}